\documentclass[11pt]{article}

\usepackage{mathtools}
\usepackage{amssymb}
\usepackage{dsfont}
\usepackage{slashed}
\usepackage{fullpage}
\usepackage{xcolor}
\usepackage[colorlinks=true,linkcolor=blue,citecolor=magenta,linktocpage=true]{hyperref}
\usepackage{titlesec}
\titleformat*{\section}{\normalsize\bfseries}
\titleformat*{\subsection}{\normalsize\bfseries}
\titleformat*{\subsubsection}{\normalsize\bfseries}
\usepackage{cite}

\DeclareMathAlphabet{\bbvar}{U}{BOONDOX-ds}{m}{n}

\makeatletter
\renewcommand{\@dotsep}{10000}
\makeatother

\newcommand{\N}{{\mathbb N}}
\newcommand{\R}{{\mathbb R}}

\newcommand{\cE}{{\mathcal E}}

\newcommand{\cO}{{\mathcal O}}

\newcommand{\cS}{{\mathcal S}}

\newcommand{\SL}{\mathrm{SL}}

\newcommand{\be}{\begin{equation}}
\newcommand{\ee}{\end{equation}}
\newcommand{\beq}{\begin{eqnarray}}
\newcommand{\eeq}{\end{eqnarray}}
\newcommand{\bes}{\begin{eqnarray}}
\newcommand{\ees}{\end{eqnarray}}

\newcommand{\mat} [2] {\left ( \begin{array}{#1}#2\end{array} \right ) }

\newcommand{\sh}{{\mathfrak{sh}}}

\renewcommand{\sl}{{\mathfrak{sl}}}

\definecolor{purple}{rgb}{0.57, 0.36, 0.75}
\definecolor{bittersweet}{rgb}{1.0, 0.44, 0.37}

\def\rd{\textrm{d}}
\def\pp{\partial}

\def\beq{\begin{eqnarray}}
\def\eeq{\end{eqnarray}}
\def\be{\begin{equation}}
\def\ee{\end{equation}}
\def\vphi{\varphi}

\def\tx{\tilde{x}}
\def\tpsi{\tilde{\psi}}

\def\vphi{\varphi}

\def\r2A{z}

\newcommand{\f}{\frac}

\def\nn{\nonumber}
\def\pp{\partial}
\numberwithin{equation}{section}

\begin{document}

\title{\Large{\textbf{\sffamily Hidden symmetry of the static response of black holes: \\Applications to Love numbers}}}
\author{\sffamily Jibril Ben Achour\;$^{1}$,  \; Etera R. Livine$^{2}$, \; Shinji Mukohyama$^{3,4}$, \;  Jean-Philippe Uzan$^{5}$}
\date{\small{\textit{$^{1}$ Arnold Sommerfeld Center for Theoretical Physics, Munich, Germany, \\
$^{2}$Univ Lyon, CNRS, ENS de Lyon, Laboratoire de Physique (LPENSL), Lyon, France\\
$^{3}$Center for Gravitational Physics, Yukawa Institute for Theoretical Physics, Kyoto, Japan\\
$^4$Kavli Institute for the Physics and Mathematics of the Universe (WPI), The University of Tokyo, Kashiwa, Chiba 277-8583, Japan\\
$^{5}$  Institut d'Astrophysique de Paris, CNRS UMR 7095, Sorbonne Universit\'es,  98 bis Boulevard Arago, 75014 Paris, France}}}

\maketitle

\begin{abstract}
We show that any static linear perturbations around Schwarzschild black holes enjoy a set of conserved charges which forms a centrally extended Schr\"{o}dinger algebra $\sh(1) = \sl(2,\mathbb{R}) \ltimes \mathcal{H}$. 
The central charge is given by the black hole mass, echoing results on black hole entropy from near-horizon diffeomorphism symmetry.
The finite symmetry transformations generated by these conserved charges correspond to Galilean and conformal transformations of the static field and of the coordinates. This new structure allows one to discuss the static response of a Schwarzschild black hole in the test field approximation from a \textit{ symmetry-based approach}. First we show that the (horizontal) symmetry protecting the vanishing of the Love numbers recently exhibited by Hui {\em et al.}, dubbed the HJPSS symmetry, coincides with one of the $\sl(2,\mathbb{R})$ generators of the Schr\"{o}dinger group. Then, it is demonstrated that the HJPSS symmetry is selected thanks to the spontaneous breaking of the full Schr\"{o}dinger symmetry at the horizon down to a simple abelian sub-group.
The latter can be understood as the symmetry protecting the regularity of the test field at the horizon. In the 4-dimensional case, this provides a symmetry protection for the vanishing of the Schwarzschild Love numbers. Our results trivially extend to the Kerr case.\footnote{YITP-22-17, IPMU22-0003}

\end{abstract}

\thispagestyle{empty}
\newpage
\setcounter{page}{1}

\hrule
\tableofcontents
\vspace{0.7cm}
\hrule

\newpage

\section{Introduction}

Tidal deformations of self-gravitating compact objects stand as key gravitational-wave observables to test General Relativity (GR) in the strong field regime. In a coalescing binary system, the deformability properties of the compact objects affect the gravitational wave signal at sub-leading post-Newtonian order through finite size effects in the pre-merger phase. These deformability properties are encoded in the so-called tidal Love numbers (TLNs) which reflect the rigidity of the system. Their measurements through gravitational wave astronomy provide rich informations, in particular to constrain the equation of state of neutron stars and the properties of compact objects beyond GR \cite{Cardoso:2017cfl, Flanagan:2007ix}. 

The relativistic theory of TLNs was introduced in Refs.~\cite{Damour:2009va, Damour:2009vw, Binnington:2009bb} and applied to a variety of compact objects, from neutron stars, black holes to more exotic self-gravitating systems \cite{Herdeiro:2020kba, Pereniguez:2021xcj, Cvetic:2021vxa, Tan:2020hog, Kim:2020dif, Brustein:2020tpg, Brustein:2021bnw}. As first shown in Refs.~\cite{Damour:2009vw, Binnington:2009bb, Fang:2005qq},  the TLNs of asymptotically flat classical and vacuum black holes accidentally vanish in 4-dimensional GR\footnote{The case of the TNLs of the Kerr black hole has been the subject of debate recently as different claims have been made. See \cite{Pani:2015hfa, Pani:2015nua, Landry:2015cva, Landry:2015zfa, Landry:2017piv, LeTiec:2020bos, LeTiec:2020spy, Poisson:2020mdi, Chia:2020yla, Charalambous:2021mea} for details.}. This property descends from demanding regularity for the tidal response of the black hole at the horizon, which forbids the existence of induced multipole moments in four dimensions. In contrast, in higher dimensions the same boundary condition still allows for induced multipoles so that black holes Love numbers are in general non-vanishing \cite{Brito:2012gw, Cardoso:2019vof, Hui:2020xxx}. Moreover, from an effective field theory point of view, the TLNs enter as suitable coupling constants in the effective action of the binary system and their vanishing thus appears as a fine tuning \cite{Porto:2016zng, Kalin:2020mvi, Kol:2011vg}.  This suggests that the vanishing of TNL in 4-dimensional GR is protected by some symmetry principle yet to be understood. Identifying this Love symmetry has thus attracting some attention recently. The role of near-horizon asymptotic symmetries has been investigated by Charalambous, Dubovsky and Ivanov~\cite{Charalambous:2021kcz}, while the role of Carollian near-horizon symmetry was discussed by Penna~\cite{Penna:2018gfx}.  

More recently, Hui, Joyce, Penco, Santoni and Solomon (HJPSS) have exhibited new symmetries, dubbed horizontal and vertical, for static linear perturbations of massless fields in Schwarz-schild and Kerr backgrounds \cite{Hui:2021vcv}. When the tidal field is slowly varying, such static test field approximation can be used to compute the Love numbers \cite{Hui:2020xxx}, making this new structure relevant to understand the vanishing of Love numbers. Concretely, the horizontal HJPSS symmetry forces the profile of the test field which is regular at the horizon to develop only a growing behavior at infinity. Hence, the static response of the black hole is absent and the associated Love number vanishes. Moreover, these symmetries allow an asymptotic observer to infer the behavior of the field at the horizon by inspecting the conservation of the charges, trading therefore a choice of boundary condition for a symmetry principle. The origin of the horizontal HJPSS symmetries remains however quite mysterious\footnote{The vertical symmetry based on the so-called ladder operators was shown to admit a geometrical origin in Ref.~\cite{Hui:2021vcv}. By mean of suitable conformal transformations, the action for the static massless test scalar field can be recast into a 3d field on a euclidean AdS space whose conformal isometries are identified with the infinitesimal generators of the HJPSS symmetry \cite{Hui:2021vcv}. This vertical symmetry will not be the focus of this work.}. What are the explicit finite transformations associated with this symmetry? Does this symmetry descend from a larger structure to be revealed? This work addresses these questions and provide a general framework to understand this new symmetry.

Since the static linear perturbations around the Schwarzschild and Kerr black holes can always be recast into a Sturm-Liouville system, we first investigate the general symmetries of this system. We stress that any system described by a Sturm-Liouville equation enjoys an infinite set of conserved charges. This property descends from the fact that for any pair of linearly independent solutions to the Sturm-Liouville equation, the associated  Wronskian $w$ is a constant of motion. The Wronskian between one such solution and the field thus provides two conserved charges ($w_1$ and $w_2$). Then, any power of these conserved charges being also conserved, one obtains an infinite tower of non-independent conserved charges which form a $w_{\infty}$ charge algebra. We then show that the truncation to linear and quadratic binomial is organized as a centrally extended one-dimensional Schr\"{o}dinger charge algebra given by the semi-direct product $\sh(1) = \sl(2,\mathbb{R}) \ltimes \mathcal{H}$. The $\sl(2,\mathbb{R})$ sector is generated by the quadratic charges while the Heisenberg algebra $\mathcal{H}$ is generated by the linear ones. There is therefore one such Schr\"{o}dinger charge algebra for any static linear perturbation on the Schwarzschild black hole. We derive explicitly the finite symmetry transformation at the level of the action, which are found to be a set of two Galilean transformations and a suitable conformal transformation of the field and the coordinate. This hidden Schr\"{o}dinger symmetry provides the general structure relevant to understand the origin of the horizontal HJPSS symmetry. 

 Indeed, applying this whole structure to the case of a static massless scalar field on the Schwarzschild black hole, we show that among the infinitesimal generators generating the $\sl(2,\mathbb{R})$ symmetry, one of them, denoted $\delta_{Q_{+}}$, coincides with the HJPSS symmetry: it leaves the growing branch invariant. However, we point that one of the other  $\sl(2,\mathbb{R})$ generators leaves instead the decaying branch invariant. From this observation, we argue that the reasoning developed in Ref.~\cite{Hui:2021vcv} has to be improved to select the symmetry protecting the growing branch (and thus the vanishing of the Love number) over the other. We show that one can discriminate between the different symmetry transformations by inspecting the on-shell values of the boundary term induced by the symmetry at the level of the action. Concretely, among the five Schr\"{o}dinger charges, only two of them generate a finite shift of the on-shell action. These two charges commuting, we conclude that the full Schr\"{o}dinger group is broken down to an abelian global symmetry group at the horizon. This residual symmetry protects the field from being divergent at the horizon.  We argue that this symmetry breaking provides the right criterion to identify the protecting symmetry associated to the purely growing behavior of the static response of the Schwarzschild black hole. This provides the second main result of this work. Let us stress that, as shown in Appendix~\ref{kerr}, the same strategy can be applied to the static response of the Kerr black hole which can also be recast in terms of a Sturm-Liouville system.
 
 While the main motivation of the present work is to provide the right framework to understand the result of Ref.~ \cite{Hui:2021vcv}, we stress that the Schr\"{o}dinger charge algebra for static fields around  a black hole provides a new structure with interesting applications for the black hole perturbation theory. We shall comment on this along this work and in the final discussion.
 
 This article is organized as follows. Section~\ref{sec1} exposes the problem and provides a brief overview of the computation of the Schwarzschild Love number in the test field approximation. We present the infinite dimensional charge algebra and its Schr\"{o}dinger sub-algebra for the Sturm-Liouville system in Section~\ref{sec31}. The finite symmetry transformations of the action are discussed in Section~\ref{sec33}. We then apply this structure to the problem in Section~\ref{sec41} where we first identify the HJPSS symmetry generator as belonging to the Schr\"{o}dinger generators. The symmetry-breaking is discussed in Section~\ref{sec42}. We conclude in Section~\ref{sec5} by a brief discussion on our results and the open directions. Finally, in Appendix~\ref{kerr}, we briefly show that the static massless scalar field on the Kerr black hole can also be recast in terms of a Sturm-Liouville equation, which shows that our results extend trivially to the rotating case. In Appendix~\ref{app}, we discuss how the static linear peturbations can be mapped onto the free particle by identifying the so-called trivializing coordinate and how it allows to understand the Schr\"{o}dinger symmetry uncovered in this work in terms of the standard Schr\"{o}dinger symmetry of the free particle.

\section{Static response of black hole: overview of the problem}

\label{sec1}

As a starting point, let us first review the standard computation of the Schwarzschild Love number in the test field approximation.
%
When the black hole is immersed in a slowly varying tidal field,
it can be considered as static and it was shown that the static test field approximation is well adapted to evaluate the Love numbers \cite{Kol:2011vg, Hui:2020xxx, Charalambous:2021kcz}. For simplicity, we shall represent the tidal field as a static massless scalar field.

Let us consider a massless scalar field living on the Schwarzschild metric given by the line element in spherical coordinates $(t,r,\theta,\phi)$,
\be
\rd s^2 = - A^2(r) \rd t^2 + \frac{\rd r^2}{A^2(r)} + r^2 \rd \Omega^2  \;, \qquad \text{with} \quad A^2(r):= 1 - \frac{r_s}{r}
\quad\textrm{and}\quad
\rd \Omega^2 =\rd\theta^{2}+\sin^{2}\theta\rd\phi^{2}
\,,
\ee
where its action is given by:
\beq
\cS[\varphi]
&=&  \frac{1}{2} \int \rd^4 x \sqrt{|g|} g^{\mu\nu} \partial_{\mu} \varphi \partial_{\nu} \varphi
\\
&=&
\frac{1}{2} \int \rd t\rd r \rd^{2}\Omega\,
\left[
-r^{2}A^{-2}(\pp_{t}\vphi)^{2}
+r^{2}A^{2}(\pp_{r}\vphi)^{2}
+(\pp_{\theta}\vphi)^{2}
+\f1{\sin^{2}\theta}(\pp_{\phi}\vphi)^{2}
\right]
\,.
\nn
\eeq
This field action consists in a kinetic term, with the time derivative, plus a potential term, depending on the spatial variations of the field.
Let us now consider a static field, thus $\pp_{t}\vphi=0$. Its reduced action consists simply the potential term of the full action, dropping both the time derivative and the time integration.
Upon integrating by part, the action of the static massless scalar field $\varphi(r,\theta,\phi)$  is then:
\be
S[\varphi] = \frac{1}{2} \int \rd r \rd \Omega^2 \left[  r^2 A^2 ( \pp_{r}\vphi)^2 -\varphi \Delta_{S_2} \varphi \right] \,,
\ee
where $\Delta_{S_2}$ is the Laplacian operator on the 2-sphere.
Static solutions, satisfying the static equation of motion
\be
\pp_{r}\left[
r^{2}A^{2}\pp_{r}\vphi
\right]
+\Delta_{S_2} \varphi 
=0
\,,
\ee
give the equilibirum configurations of the dynamical model.

Decomposing the field in spherical harmonics, 
\be
\vphi(r,\theta, \phi) = \sum_{\ell\in\N}\sum_{m=-\ell}^{+\ell} \vphi_{\ell, m} (r) Y^{m}_{\ell} (\theta, \phi) \,,
\ee
we diagonalize the spherical Lagrangian and obtain one-dimensional differential equations for the radial factors:
\be
\r2A\pp_{r}^{2}\vphi_{\ell, m}+\pp_{r}\r2A\pp_{r}\vphi_{\ell, m}-\ell(\ell+1)\vphi_{\ell, m}=0\,,
\ee
where we have introduced the notation
\be
\r2A := r^2 A^2(r) = r(r-r_s)
\,.
\ee
This function $\r2A$  matches the quantity $\Delta$ in Ref.~\cite{Hui:2021vcv}, while avoiding potential confusion with the Laplacian and finite  variations.

\subsection{Legendre solutions and Love number}

In order to solve these equations of motion, putting the label $m$ aside, we notice that a simple change of variable allows to write them as Legendre differential equations:
\be
\vphi_{\ell}(r)=f_{\ell}(x) \quad\textrm{with}\,\,\;\;\;x=\f{2r}{r_{s}}-1
\qquad\Rightarrow\quad
(1-x^{2})f_{\ell}''-2xf_{\ell}'+\ell(\ell+1)f_{\ell}=0
\,.
\ee
It is well-known that such a second order differential equation admits two independent solutions, the Legendre functions of the first and second kinds  $P_{\ell}(x)$ and $Q_{\ell}(x)$. These functions are usually defined on the interval $x\in]-1,+1[$, but their definition is straightforwardly extendable to the half-line $x\in]+1,+\infty[$ corresponding to the exterior of the black hole $r\in \;  ]r_s, +\infty [$.
The $P_{\ell}$'s are the Legendre polynomials, the $Q_{\ell}$'s differ by a logarithmic factor. The $P_{\ell}$'s take  finite values at the horizon $r=r_{s}$ and diverge at $r\rightarrow +\infty $, while the $Q_{\ell}$'s have the opposite behavior, they diverge at the horizon and asymptotically vanish as $r$ grows to infinity.
\footnotetext{ 
The Legendre functions $P_\ell(x)$  are polynomials  when $\ell$ is an integer, and thus are well-defined and real-valued on the whole real line ${\mathbb R}$. They only diverge at infinity. Concerning $Q_\ell(x)$, they are  usually defined on $]-1,+1[$, corresponding to $r/r_s\in ]0,1[$. Adapting their definition to the exterior interval is straightforward. Starting from $\ell=0$, we slightly modify the definition of the $Q_{0}$,
\be
Q_0(x)= \frac{1}{2}\ln\frac{1+x}{1-x}, \,\,\forall x\in ]-1,+1[
\quad{\rm and}\quad  Q_0(x)= \frac{1}{2}\ln\frac{x+1}{x-1}, \,\,\forall x\in ]+1,+\infty[\,,
\ee
then construct the higher angular momenta solutions $Q_\ell$ from the Bonnet recursion as
\be
Q_1(x)= x Q_0 -1
\,,\qquad
Q_\ell(x)= \frac{2\ell-1}{\ell}Q_{\ell-1} - \frac{\ell-1}{\ell} Q_{-2} \quad\textrm{for}\,\,\ell\ge2
\,.
\nn
\ee
They can also be expressed in terms of hypergeometric functions defined on $]1,+\infty[$. as
\be
Q_\ell(z)=\frac{\Gamma(\ell+1)\Gamma\left(\frac{1}{2}\right)}{2^{\ell+1}\Gamma\left(\ell+\frac{3}{2}\right)} x^{-(\ell+1)}
{}_2F_1\left(\frac{\ell+2}{2},\frac{\ell+1}{2};\ell+\frac{3}{2}; z^{-2} \right)
\,.
\nn
\ee
}
%
%
A general solution is thus a superposition of the two Legendre functions,
\be
\vphi_{\ell}(r) = c_1 \left(\vphi_{\ell}\right)_1+ c_2 \left(\vphi_{\ell}\right)_2  \;, \qquad \text{with} \qquad 
\left\{
    \begin{array}{ll}
        (\vphi_{\ell})_1 & =  P_{\ell} \left( \frac{2r}{r_s} - 1\right) \\
        (\vphi_{\ell})_2 & = Q_{\ell} \left( \frac{2r}{r_s} - 1\right)
    \end{array}
\right. \,,
\ee
where the two branches and their asymptotic are given by:
\begin{align}
&(\vphi_{\ell})_1(r)\underset{r\rightarrow r_{s}^{+}} {\sim}
1
\,,\qquad
\qquad \qquad \qquad \;\;\;\;\;\; (\vphi_{\ell})_1(r)\underset{r\rightarrow +\infty}{\sim} \frac{2^{2\ell} \Gamma(\ell+ 1/2)}{\sqrt{\pi} \ell !} \left( \frac{r}{r_s}\right)^{\ell}
\,, \\
&(\vphi_{\ell})_2(r)\underset{r\rightarrow r_{s}^{+}}{\sim} - \frac{1}{2 \ell !} \log{ \left[ 2 \left( \frac{r}{r_s} -1 \right)\right]}
\,,\qquad
(\vphi_{\ell})_2(r)\underset{r\rightarrow +\infty}{\sim} \frac{ 2^{2\ell}\sqrt{\pi}}{\Gamma(\ell+3/2)} \left( \frac{r}{r_s}\right)^{-\ell-1}
\,.
\end{align}
The coefficients $(c_1, c_2)$ are  a priori arbitrary constants determining a general solution. But due to the divergent behavior of the two chosen solutions $(\vphi_{\ell})_1$ and $(\vphi_{\ell})_2$ either at the horizon or at spatial infinity, those coefficients will be fixed by imposing suitable boundary conditions.
Since the   $Q_{\ell}$ solution
 diverges at the horizon because of the logarithmic factor,
and we require that the tidal field be regular at the horizon, this imposes that $c_2=0$ and that the only admissible solution is the first branch
\be
\varphi_{\ell}(r) = P_{\ell} \left( \frac{2r}{r_s} - 1\right) \;,
\qquad \text{hence} \qquad
\varphi_{\ell}(r) \underset{r \gg r_s}{\propto} r^{\ell}
\,.
\ee
Therefore, the profile of the static test scalar field does not exhibit any tail. This means that the Love number  vanishes in this case. Indeed the Love number is
defined as the ratio between the coefficient of the leading order of the tail (i.e. the induced $r^{\ell+1}$ tail at infinity arising from the $Q_\ell$ mode) and the coefficient of the leading order to the growing behavior (i.e. the $r^\ell$ tidal mode associated with $P_\ell$).
The vanishing of the coefficient $c_{2}$ is thus equivalent to the vanishing of the Love number.
The same is true for spin-1 and spin-2 static linear perturbations around the Schwarzschild black hole in four dimensional GR. However, in higher dimensions or in modified gravity, this boundary condition does not prevent the field from exhibiting a tail and thus the Love number are in general non-vanishing \cite{Hui:2020xxx}.

From that perspective, the vanishing of the Love numbers in 4-dimensional GR is a surprising property. The goal of the following sections is to revisit this problem from the point of view of its hidden symmetries to study whether it can be protected. While the problem is straightforward to solve by using suitable boundary conditions, i.e. $c_2=0$, the symmetry of the system allows one to trade this physical boundary condition into a symmetry argument, following the idea first advocated in Ref.~\cite{Hui:2021vcv}. Identifying the conserved charges and the explicit symmetry transformation which fully dictate the profile of the static test scalar field thus provides a way to characterize the static response of the Schwarzschild black hole and the vanishing of its Love numbers from a purely symmetry-based approach.

\subsection{Sturm-Louiville problem}
In order to investigate the general structures and symmetries of the model in a systematic fashion, it is convenient to reformulate it as a Sturm-Louiville problem, i.e. as a free massless 1D field evolving in a non-trivial potential. This reformulation underlines the universality of the approach and of our results.

To this purpose, we introduce a rescaled scalar field $\psi$ 
\be
\label{rescaling}
\psi := \sqrt{\r2A}  \varphi
\,.
\ee
This rescaling removes the friction term in the equation of motion. Using the  prime notation for radial derivatives, $\psi'\equiv \pp_{r}\psi$, the static action becomes:
\be
S[\psi] = \frac{1}{2} \int \rd r \rd \Omega^2  \left[  (\psi')^2 - \frac{\mu^2}{\r2A^2} \psi^2 - \frac{1}{\r2A} \psi \Delta_{S_2} \psi \right]
\,,
\ee
where $\mu^{2} = r^2_s/4$ is the horizon area up to a numerical factor. 
Decomposing the field in spherical harmonics\footnotemark, $\psi(r,\theta, \phi) = \sum_{\ell, m} \psi_{\ell, m} (r) Y^{m}_{\ell} (\theta, \phi)$,
\footnotetext{
Even if the scalar field is assumed to be real, its decomposition on spherical harmonics is not. Due to the reality conditions satisfied by the spherical harmonics, $\overline{Y}_{\ell}^{m}=(-1)^{m}Y_{\ell}^{-m}$, the modes of a real field  satisfy $\overline{\psi}_{\ell,m}=(-1)^{m}\psi_{\ell,-m}$.
}
%
the static action becomes a sum of decoupled harmonic oscillators corresponding to each multipole
\be
\label{acc}
S[\psi] = 4\pi \sum_{\ell, m} \frac{1}{2} \int \rd r \left\{ | \psi'_{\ell, m}|^2 - V_{\ell}(r) |\psi_{\ell, m} |^2\right\} \,,
\ee
where the potentials are given by
\be\label{e.Vl}
V_{\ell} (r) := \frac{\mu^2}{\r2A^2} - \frac{\ell(\ell+1)}{\r2A}
\,,
\ee
and depends only on the angular momentum label $\ell$ thanks to spherical symmetry. One can thus investigate the dynamics of each multipole independently. The equation of motion for a field mode $\psi_{\ell, m}$ is given by a Sturm-Liouville equation,
\be
\label{eom}
\cE_{\ell} \psi_{\ell, m} := \left[ \rd^2_r + V_{\ell}(r) \right]  \psi_{\ell, m} =0
\,.
\ee
Such a second order linear differential equations admits two independent solutions. We recover the Legendre solutions obtained for the original unrescaled scalar field $\vphi=\psi/\sqrt{z}$:
\be
\label{sol}
\psi_{\ell}(r) = c_1 \left(\psi_{\ell}\right)_1+ c_2 \left(\psi_{\ell}\right)_2   \qquad \text{with} \qquad \left\{
    \begin{array}{ll}
        (\psi_{\ell})_1 = \sqrt{\r2A} P_{\ell} \left( \frac{2r}{r_s} - 1\right)\,, &  \\
         (\psi_{\ell})_2 = \sqrt{\r2A} Q_{\ell} \left( \frac{2r}{r_s} - 1\right) \,.& 
    \end{array}
\right.
\ee
Before going further, let us point that the same strategy can be applied the static massless scalar field on the Kerr black hole, such that it can also be recast in terms of a Sturm-Liouville equation with potential (\ref{pot}). See Appendix~\ref{kerr} for details.
\subsection{Ladder operators for angular momenta}

It turns out that the profile of an arbitrary $\ell$-multipole can be fully deduced from the profile of the zero mode \cite{Hui:2021vcv}. This will allow us to focus on the simplest sector with zero angular momentum when discuss the symmetries of the system in the next section. This is realized through the existence of ladder operators allowing to switch between different values of the angular momentum. This is especially clear in terms of Sturm-Louiville operators.

Let us consider the rescaled equation of motion $E_{\ell} \psi_{\ell}= \r2A \cE_{\ell} \psi_{\ell} \simeq 0$ and introduce pairs of first order differential operators:
\be
\label{ladder}
D_{\ell}^{+}=\r2A \pp_{r}+\f{(\ell-1)}2\r2A'
\,,\qquad
D_{\ell}^{-}=\r2A\pp_{r}-\f{(\ell+2)}2\r2A'
\,.
\ee
They satisfy the following intertwining commutation relations
\be
D_{\ell}^{-}E_{\ell+1}=E_{\ell}D_{\ell}^{-}
\,,\qquad
D_{\ell}^{+}E_{\ell-1}=E_{\ell}D_{\ell}^{+} 
\,,
\ee
such that they act as ladder operators between the space of solutions with different angular momenta:
\be
E_{\ell+1}\psi=0\,\,\Rightarrow\,\, E_{\ell}(D_{\ell}^{-}\psi)=0
\,,\qquad
E_{\ell-1}\psi=0\,\,\Rightarrow\,\, E_{\ell}(D_{\ell}^{+}\psi)=0
\,.
\ee
Thus, starting from the $\ell=0$ solution, i.e. $E_0 \psi_0 = 0$, similar to highest weight vectors, one can construct a solution in the $\ell$ sector as
\be
\psi_{\ell} = D^{+}_{\ell-1} ..... D^{+}_{2} D^{+}_1 \psi_0
\,.
\ee
On the contrary, the operator $D^{-}_{\ell}$ allows one to climb down the ladder.
 Let us point that the ladder operators satisfy themselves ladder relations, showing that they can be generated from the lowest angular momenta operator b a simple rescaling:
\be
\sqrt{z}D^{+}_{\ell+1}=D^{+}_{\ell}\sqrt{z}
\qquad\Rightarrow\quad
z^{\f{\ell-1}2}D^{+}_{\ell}=D^{+}_{1}z^{\f{\ell-1}2}
\,,
\ee
\be
\sqrt{z}D^{-}_{\ell-1}=D^{-}_{\ell}\sqrt{z}
\qquad\Rightarrow\quad
D^{-}_{\ell}z^{\f{\ell}2}=z^{\f{\ell}2}D^{-}_{0}
\,.
\ee
Although we will not focus on this structure in the present work, we stress that these ladder operators generate a symmetry of the test field action dubbed vertical symmetry in Ref.~\cite{Hui:2021vcv}. Finally, let us point out that a similar ladder structure can be constructed which connect the solutions for different spins. See Refs.~\cite{Hui:2021vcv, Cardoso:2017qmj} for details.
This concludes the overview of the problem. We now turn to the original part of the present work and move to the general symmetries of the system.
 
\section{Schr\"{o}dinger symmetry for Sturm-Liouville systems}
\label{sec3}

Any static linear perturbations around a Schwarzschild and Kerr black holes follows an equation of motion which can be recast into a Sturm-Liouville operator of the form (\ref{eom}). In the following section, we point out the relation between the conservation of the Wronskian of pairs of solutions to the e.o.m. and the existence of an infinite tower of conserved charges for the system. In particular, we show that within this infinite dimensional charge algebra, one can identify a finite dimensional subalgebra generating a Schr\"{o}dinger global symmetry of the Sturm-Liouville system for which we derive the explicit finite symmetry transformations. 

Let us consider a one-dimensional field $\psi$ depending on one coordinate $x$, which can interpret as the ``time'' along which we describe the evolution of the field. We define the action as
\be
\label{ac}
S[\psi] = \int \rd x\,L[x,\psi]\,,
\qquad\textrm{with}\quad
L[x,\psi]
=
\frac{1}{2}  \Big{[} (\psi')^2 - V(x) \psi^2 \Big{]}
\,,
\ee
such that its e.o.m. is given by a Sturm-Liouville operator,
\be
\label{eomm}
\cE \psi := \left[\rd^2_x + V(x) \right] \psi \simeq 0 \;,
\ee
where $\simeq$ denotes on shell from now on.
A general solution reads $\psi \simeq c_1 \psi_1 + c_2 \psi_2$ in terms of  two linearly independent solutions $(\psi_1, \psi_2)$.
The two coefficients $(c_1, c_2)$ are arbitrary constants to be fixed by suitable boundary conditions (at infinity and at the poles of the potential).
The Hamiltonian formulation of the system is given by
\be
p = \f{\delta L}{\delta \psi'}=\psi'
\;, \qquad
H = p\psi'-L=\frac{1}{2} \left( p^2 + V(x) \psi^2 \right) \;, \qquad  \left|
    \begin{array}{lclcl}
        \psi' &=& \{\psi, H \} &=&  p      \,,\\
        p' &=& \{ p, H\} &=& - V \psi     \,, 
         \end{array}
    \right.
\ee
where the Poisson bracket is defined by the canonical pair $\{ \psi, p\} =1$.

Let us now move on to identifying the  symmetries of the system. 
Since the potential $V$ is explicitly $x$-dependent, the Hamiltonian is not conserved (with respect to the evolution in the $x$-coordinate). So we have to identify other conserved quantities and understand the symmetry transformations that they generate by Noether's theorem.

\subsection{Conserved Wronskian, Schr\"{o}dinger algebra and central charge}
\label{sec31}

A special property of the Sturm-Louiville system is that the Wronskian of two arbitrary solutions of the equation of motion is a constant of motion. For instance, for the  two linearly independent solutions $( \psi_1, \psi_2 )$, we compute the derivative of the Wronskian $w[\psi_1, \psi_2] =  (\psi_1 \psi'_2 - \psi_2 \psi'_1 )$:
\be
\rd_{x }w[\psi_1, \psi_2] =  (\psi_1 \psi''_2 - \psi_2 \psi''_1 ) \simeq V(x) \left( -  \psi_1 \psi_2 + \psi_2 \psi_1 \right) =0 \,.
\ee
In the following, we shall use the shorthand notation $w:= w[\psi_1, \psi_2]$ which is a number.

Thus, for an arbitrary classical solution,  one can construct two conserved charges given by its Wronskian with each of the two chosen solutions:
\be
\label{cons}
w_1:= w[\psi_1, \psi] = \psi_1 \psi' - \psi \psi_1'\;, \qquad w_2 := w[\psi_2, \psi]=  \psi_2 \psi' - \psi \psi_2'
\,.
\ee
These charges turn out to be conjugate variables,
\be
\{ w_1, w_2\} =w
\,.
\ee
One can go further and construct an infinite tower of conserved charges for this system, consisting in all the polynomials $w_{1}^{n_{1}}w_{2}^{n_{2}}$. They form a $w_{\infty}$ charge algebra:
\be
\label{infch}
\{w_{1}^{n_{1}}w_{2}^{n_{2}},w_{1}^{m_{1}}w_{2}^{m_{2}}\}
=
w(n_{1}m_{2}-m_{1}n_{2})\, w_{1}^{n_{1}+m_{1}-1}w_{2}^{n_{2}+m_{2}-1}
\,,
\ee
%
%
A natural question is then whether one can identify a closed finite dimensional charge algebra within this infinite dimensional Lie algebra.
We can indeed consider the truncation to linear and quadratic binomials.
Denoting the linear and quadratic charges as
\begin{align}
\label{Sch}
 \left|
    \begin{array}{l}
         Y_{+}  = w_1 \,,  \\
         Y_{-}  = w_2 \,, 
    \end{array}
\right. 
\qquad \qquad 
\left|
    \begin{array}{l}
         Q_{+}= w^2_1/2 \,,  \\
         Q_{-} = w^2_2/2 \,,  \\
         Q_{0} = w_1 w_2/2 \,.
    \end{array}
\right.
\end{align}
we compute the commutators of these five constants of motion:
\begin{align}
 \{ Q_{+}, Q_{-}\} = 2 w Q_0 \;, \qquad &\{ Q_{0}, Q_{+}\} = - w Q_{+} \;, \qquad \{ Q_{0}, Q_{-}\} = w Q_{-}
\,,  \\
 \{ Q_0, Y_{\pm}\} = \mp \frac{w}{2} Y_{\pm} \;, \qquad &  \{ Q_{+}, Y_{-}\} = w Y_{+} \;,\quad \qquad \{ Q_{-}, Y_{+}\} = - w Y_{-}
\,, \\
&\{ Y_{+}, Y_{-}\} = w\,.&
\label{cext}
\end{align}
The first line corresponds to an $\sl(2,\mathbb{R})$ algebra (with vanishing quadratic Casimir)
while the last bracket defines a  Heisenberg algebra $\mathcal{H}$.
We recognize the one-dimensional Schr\"{o}dinger algebra as the semi-direct product $\sh(1) = \sl(2,\mathbb{R}) \ltimes \mathcal{H}$, where the central charge is given by the conserved Wronskian $w$. To our knowledge, this is the largest finite dimensional closed algebra one can construct. See Refs.~\cite{Sym1, Carinena:2017svz} for complementary investigations on the symmetries of specific second order ordinary differential equations and \cite{Cariglia:2018mos, Cariglia:2016oft, Zhao:2021tsz} for a discussion on the dynamical symmetry of time-dependent systems.

For a given solution $\psi \simeq c_1 \psi_1 + c_2 \psi_2$, the Wronskian are $w_1(\psi)=c_2w$ and $w_2(\psi)=-c_1w$, so that the on-shell values of the Schr\"{o}dinger charges are given by
\begin{align}
\label{onshell}
\left|
    \begin{array}{l}
         Y_{+}  \simeq c_2 w \,,  \\
         Y_{-}  \simeq  -c_1 w \,,
    \end{array}
\right. 
\quad\qquad 
\left|
    \begin{array}{l}
         Q_{+} \simeq c^2_2w^2/2 \,, \\
         Q_{-} \simeq c^2_1 w^2/2 \,,  \\
         Q_{0}\;  \simeq -  c_1 c_2 w^2/2\,. 
    \end{array}
\right.
\end{align}
Now, those charges are supposed to generate symmetries by Noether's theorem. Let us thus derive the action of the resulting finite symmetry transformations on the solution space.

\subsection{Infinitesimal action on fields}
\label{sec32}

The infinitesimal action of each charge on the field  is given by the Poisson bracket $\delta \psi = \{ w^{n_1}_1 w^{n_2}_2, \psi\}$.
First, the linear charges induce the following field variations:
\be
\delta_{Y_{+}} \psi  = \{ Y_{+}, \psi\} = -  \psi_1\;, \qquad \delta_{Y_{-}} \psi  = \{ Y_{-}, \psi\} = - \psi_2.
\ee
Those Poisson brackets are immediate to exponentiate and leads to finite translations of the field $\psi$ along solutions to the equations of motion. These are Galilean transformations and we will check in the next section that they are indeed symmetries of the system. While these transformations may seem trivial at first glance, they however encode relevant information through the central extension of the Heisenberg algebra they form, i.e. Eq~(\ref{cext}). As we shall see in the next section, this central extension corresponds to the mass of the black hole for the zero mode static perturbation around the Schwarzschild black hole.

Now, the $\sl(2,\mathbb{R})$ charges induce the following field variations:
\begin{align}
\label{trans1}
 \delta_{Q_{+}} \psi  & = \{ Q_{+}, \psi\} = -w_{1} \psi_1 \;, \\
 \label{trans2}
  \delta_{Q_{-}} \psi  & = \{ Q_{-}, \psi\} =  - w_{2} \psi_2\;, \\
   \delta_{Q_0} \psi  & = \{ Q_{0}, \psi\} = -\frac{1}{2} (w_{1} \psi_{2}+ w_{2} \psi_{1})
   \,.
\end{align}
Notice that the trnasformations (\ref{trans1}) and (\ref{trans2}) are linear in $\psi$ as one can see from the definition Eq.~(\ref{cons}). Although these variations are not straightforward to interpret, they considerably simplify on-shell. Indeed, for an arbitrary solution $\psi \simeq c_1 \psi_1 + c_2 \psi_2$, the $\sl(2,\mathbb{R})$ charges act as follows:
\begin{align}
 \delta_{Q_{+}} \psi  &\simeq -c_2 w \psi_1 \;, \\
  \delta_{Q_{-}} \psi  &\simeq  + c_1 w \psi_2\;, \\
   \delta_{Q_0} \psi  & \simeq \frac{1}{2} w (c_2 \psi_2 - c_1 \psi_1)
   \,.
\end{align}
The first charges $Q_{\pm}$ shift the solution along one given branch while the charge $Q_0$ squeezes the two branches. 
Those infinitesimal variations clearly map solutions of the e.o.m. to other solutions of the e.o.m. We can thus describe the action on the solution space by looking at how the coefficients $c_{1}$ and $c_{2}$ are affected. 
We can indeed represent these as standard 2$\times$2 matrices acting on the real 2-vector $(c_{1},c_{2})$ in the $(\psi_1, \psi_2)$-basis:
\begin{align}
 \delta_{Q_{+}} \mat{c}{c_{1}\\c_{2}}
& =
-w\mat{cc}{0 & +1\\ 0 & 0}\mat{c}{c_{1}\\c_{2}}
\,,\\
  \delta_{Q_{-}} \mat{c}{c_{1}\\c_{2}}
& =
-w\mat{cc}{0 & 0\\ -1 & 0}\mat{c}{c_{1}\\c_{2}}
\,,\\
  \delta_{Q_{0}} \mat{c}{c_{1}\\c_{2}}
& =
-\frac{w}{2}\mat{cc}{1 & 0\\ 0 & -1}\mat{c}{c_{1}\\c_{2}}
\,.
\end{align}
We recognize a basis of the $\sl(2,\R)$ Lie algebra, which generates the $\SL(2,\R)$ group. 
The variation along $ \delta_{Q_{+}}$ exponentiates to upper triangular matrices, $  \delta_{Q_{-}} $ exponentiates to lower translations, while $  \delta_{Q_{0}} $ generates dilatations.
This fully describes the action of the $\SL(2,\R)$ symmetry transformations on the space of classical solutions to the field equation.
We describe below how the $\SL(2,\R)$ group transformations act on arbitrary fields: they form the subgroup of the Virasoro group  of conformal transformations which preserve the potential.

\subsection{Finite conformal transformations}
\label{sec33}

In this section, we identify the finite symmetry transformations generated by the Schr\"odinger charges and show that these are indeed the corresponding Noether charges.

Let us start with  the following Galilean transformation
\begin{align}
\label{sym2}
x \rightarrow \tilde{x} &=x \,, \\
\label{sym22}
\psi \rightarrow \tilde{\psi}(\tilde{x}) &=  \psi(x) + \chi (x) \,,
\end{align}
where $\chi(x)$ is a priori an arbitrary function. We are interested in understanding when such a field translation is a symmetry of the system.
The Sturm-Liouiville action \eqref{ac} transforms as
\be
\Delta S = {S}[\tilde{x}, \tilde{\psi}] - S[x, \psi]  = \int \rd x \left\{ \frac{\rd}{\rd x} \left( \chi \chi' + 2 \psi \chi'\right) - (\chi + 2 \psi) (\chi'' + V \chi)\right\}
\, .
\ee
It is a Noether symmetry when the action variation is solely a boundary term, i.e. the integral of a total derivative.  This happens if and only if $(\chi'' + V \chi)$ vanishes, that is when the translation parameter $\chi$ is a solution of the e.o.m. In that case, $\chi$ can be decomposed onto the chosen independent solutions,
\be
\label{cond}
\chi'' + V(x) \chi =0  \qquad \Leftrightarrow \;\qquad \chi = \eta_{+} \psi_1 + \eta_{-} \psi_2
\ee
where $\eta_{\pm}$ are real constants labelling this global symmetry. This means that this is a two-parameter symmetry of the system. We would like to stress that these are proper  {\em off-shell symmetries}: while the symmetry parameters are constrained by the e.o.m., the field $\psi$ is arbitrary and is not assumed to satisfy the e.o.m. in any measure.
%
%
The Noether charges generating this symmetry are easily computed from the total derivative term of the Lagrangian variation and read
\be
Y [\chi, \psi]= \delta\psi\f{\delta L}{\delta \psi'} -  \psi \chi'= \chi \psi' - \chi' \psi  + \cO(\chi)
\,,\qquad
\left|
    \begin{array}{lcl}
\chi=\psi_{1}&\Rightarrow&  Y_{+}[\psi_1, \psi] =  w_1\,,  \\
\chi=\psi_{2}&\Rightarrow & Y_{-}[\psi_2, \psi] =  w_2 \,. 
    \end{array}
\right.
\ee
Notice that we have neglected the term $\chi \chi'$ which is second order in $\chi$ in the expression of the charges.
As expected, the Wronskians $w_1$ and $w_2$ thus generate the Galilean symmetry transformations of the Sturm-Louiville system. 

\bigskip

Now, let us analyse the symmetries generated by the quadratic charges given by the squared Wronskians. We introduce  conformal reparametrizations parametrized by an a priori arbitrary function $f(x)$:
\be
\label{sym1}
\begin{array}{rcrcl}
x &\mapsto &\tilde{x} &= &f(x) \,,
\\
\psi(x) &\mapsto& \tilde{\psi}(\tilde{x}) &= &f'(x)^{1/2} \psi(x)\,,
\end{array}
\ee
where the field $\psi(x)$ transforms as a primary field with conformal weight $\f12$.
Under such finite transformation, the action of our system transforms as
\be
S[\tx,\tpsi(\tx)]
=
\f12\int \rd x\,
\left[
(\rd_{x}\psi)^{2}
 -\left( \frac{1}{2}\text{Sch}[f] + (f')^2(V \circ f)\right) \psi^2
+ \frac{1}{2} \frac{\rd }{\rd x} \left( \frac{f''}{f'} \psi^2\right)
\right]
\,.
\ee
Up to a total derivative (i.e. a boundary term), this is again a Sturm-Louiville system with a modified potential:
\be
\tilde{V}=(f')^2(V \circ f)+\frac{1}{2}\text{Sch}[f]
\qquad\textrm{with}\quad
 \text{Sch}[f] = \frac{f'''}{f'} - \frac{3}{2} \left( \frac{f''}{f'}\right)^2
\,,
\ee
where $ \text{Sch}[f]$ is the Schwarzian derivative\footnotemark{} of the function $f(x)$.
\footnotetext{
Two key properties of the Schwarzian are given by:
\be
\text{Sch}[f] =0 \,\, \Leftrightarrow \,\, f(x) = \frac{a x+b}{cx+d}  \,,\quad\textrm{and}\quad
\text{Sch}[f_1\circ f_2] = \text{Sch}[f_2] + (f'_2)^2 \left(  \text{Sch}[f_1] \circ f_2\right)  \,. \nn
\ee
The first property identifies M\"obius transformations as the kernel of the Schwarzian derivative, while the second equality is the cocycle property resulting in the consistent deformation of the one-dimensional diffeomorphisms into the Virasoro group.
}
This is exactly the Virasoro group action on Sturm-Louiville operators (e.g. \cite{ovsienko2006}), where we recognize the non-trivial Schwarzian cocycle. This cocycle term signals a deformation of the action of one-dimensional diffeomorphisms on the potential.
Here we are not interested in a full description of the Virasoro group and its orbits, and we would like to focus on the symmetries of the system. These  conformal reparametrizations are symmetries when the action variation is a mere boundary term,
\be
\Delta S = \f12\int \rd x \left\{ \frac{1}{2}\frac{\rd }{\rd x} \left( \frac{f''}{f'} \psi^2\right) - \left[ \frac{1}{2}\text{Sch}[f] + (f')^2(V \circ f) -  V\right] \psi^2 \right\}
\,,
\ee
which is equivalent to the potential remaining unchanged, that is if the transformation parameter $f$ satisfies the following functional condition:
\be
\label{cond0}
 \text{Sch}[f] = 2 V - 2  (f')^2(V \circ f)
 \,.
\ee
The solution to this third order differential equation is rather complicated to write in a closed form. However, notice that depending on the form of the potential $V$, one can sometimes obtain an explicit solution for the finite transformation $f(x)$. This will be the case for the application to black hole we shall present later in Section~\ref{sec5}.


At this stage, it is not clear that solutions to \eqref{cond0} form a $\SL(2,\R)$ group.  The simplest route to show that this is indeed the case is to identify a ``trivializing map'' $f$ such that the corresponding conformal reparametrization yields a vanishing potential. This allows to map the system with a potential onto a free field. We describe this ``conformal bridge'' method in more details in Appendix~\ref{app}. For the moment, let us briefly explain its construction. Let us call $f$ such a trivializing map. This map transforms the original coordinate and field $(x, \psi)$ onto a new system $(F, \Phi)$ such that $\Phi$ satisfies the trivial field equation $\rd_{F}^{2}\Phi=0$. It turns out that the new trivializing coordinate $F$ is simply given by the ratio of two linearly independent solutions, for instance, $F=\psi_{2}/\psi_{1}$. Once we work with the free system, we readily check that it is invariant under conformal reparametrization given by  standard M\"{o}bius transformation in the coordinate:
\be
\label{mob}
\begin{array}{rcrcl}
F &\mapsto &\tilde{F} &= &{\displaystyle M(F)=\f{(\alpha F+\beta)}{(\gamma F+\delta)} }
\quad\textrm{with}\,\,(\alpha\delta-\beta\gamma)=1 \,,
\vspace*{2mm}\\
\Phi(F) &\mapsto& \tilde{\Phi}(\tilde{F}) &= &{\displaystyle M'(F)^{\f12} \Phi(F)}
\,,
\vspace*{2mm}\\
\rd_{F}^{2}\Phi &\mapsto& \rd_{\tilde{F}}^{2}\tilde{\Phi}&=&(\gamma F +\delta)^{3}\,\rd_{F}^{2}\Phi 
\,.
\end{array}
\ee
These M\"{o}bius transformations form a $\SL(2,\R)$ group. We can cross the bridge back to the system with a potential: the symmetry is given by conformal reparametrizations $F^{-1}\circ M \circ F$, which satisfy the condition \eqref{cond0} and clearly form a $\SL(2,\R)$ group. This $\SL(2,\R)$ symmetry can be understood directly at the level of the existence of the trivializing coordinate $F$. Indeed we have an obvious freedom in choosing the two linearly independent solutions used to define $F$. We can switch the ratio $\psi_{2}/\psi_{1}$ with any ratio $(\alpha\psi_{2}+\beta\psi_{1})/(\gamma\psi_{2}+\delta\psi_{1})$ as long as $(\alpha\delta-\beta\gamma)\ne 0$.
At this point, we can anticipate on the symmetry breaking mechanism that we will uncover in the context of the black hole tidal modes in the next section. Indeed, when the potential admits poles, it leads to distinct boundary conditions and asymptotic behaviors for the two solutions $\psi_{1}$ and  $\psi_{2}$, then we lose the freedom to consider arbitrary linear combination of those two solutions. This leads to a spontaneous symmetry breaking from the whole $\SL(2,\R)$ group down to an abelian $\R$ subgroup.

\medskip

Coming back to the symmetry (\ref{sym1}) with the condition (\ref{cond0}), a little work is required to show that the charges generating these symmetries coincide with the squared Wronskians. Consider the infinitesimal conformal reparametrizations, with $f(x)= x + \xi(x)$. At first order in $\xi$, the field variation reads
\be
\delta\psi=\tpsi(x)-\psi(x)=\xi\psi' -\f12\xi'\psi\,,
\ee
and the condition (\ref{cond0}) becomes
\be
\label{lam}
\xi''' + 2V' \xi + 4 V \xi'  =0
\,.
\ee
An amazing property is that  solutions to this third order differential equation can be constructed from solutions to the Sturm-Louiville e.o.m. Indeed if $\chi$ is a solution to $\chi'' + V(x) \chi =0$, then $\xi=\chi^{2}$ is a solution to the symmetry condition \eqref{lam}. This means that one can construct suitable symmetry parameters from the two linearly independent solutions $(\psi_1, \psi_2)$:
\be
\label{lam2}
\xi(x) = \alpha_{+} \psi^2_1 + \alpha_{-} \psi^2_2 + \alpha_0 \psi_1 \psi_2
\,,
\ee
in terms of three arbitrary coefficients $(\alpha_{+}, \alpha_{-}, \alpha_0)$. Since this defines a three-dimensional space of solutions, this provide the whole solution space for the  third order differential equation.
%
Applying the Noether's theorem, we compute the associated Noether charges from the Lagrangian variation,
\be
\label{ch}
Q_{\xi_i} [ \psi] = \frac{1}{4} \xi''_i \psi^2 - \frac{1}{2} \xi'_i \psi \psi' + \frac{1}{2} \xi_i \left( (\psi')^2 + V \psi^2\right)
\,.
\ee
We check that these are indeed conserved charges:
\be
Q'_{\xi_i} = \f14 \left( \xi_i''' + 2V' \xi_i + 4 V \xi_i'  \right)\psi^{2}+\,\left(  \xi_{i} \psi'-\frac{1}{2} \xi'_i \psi \right)\cE[\psi] \simeq 0 \,,
\ee
where the first term vanishes because of the condition (\ref{lam}) while the second term vanishes on-shell when $\psi$ is a solution to the e.o.m. $\cE[\psi] \simeq0$. Now, a straightforward computation shows that the three charges can be compactly written as the squared Wronskians:
\be
\left|
    \begin{array}{rcl}
         Q_{+}[\psi^2_1, \psi]&=& w^2_1/2\,,   \\
         Q_{-}[\psi^2_2, \psi] &=& w^2_2 /2\,,  \\
         Q_{0} [\psi_1 \psi_2, \psi]&=& w_1 w_2/2\,.
    \end{array}
\right. 
\ee
in agreement with the result of Section~\ref{sec31}. At this stage, we have closed the circle and identified the finite transformations generated by the conserved charges discussed in Section~\ref{sec3}.
Before going further, let us stress that the symmetry discussed here  applies to any static field on a spherically symmetry background. In particular, the action (\ref{ac}) can also account for massive fields as well as higher spin field. 

We can now apply this structure to the problem of the static response of the Schwarzschild black hole. As we shall see, the Schr\"{o}dinger symmetry provides the right framework to identify the origin of the horizontal HJPSS symmetry introduced in Ref.~\cite{Hui:2021vcv} for static massless test fields.

\section{Application to the static response of Schwarzschild black holes}

Consider a static massless test scalar field on a Schwarzschild black hole. Each multipole admits the following action
\be
\label{acmulti}
S[\psi_{\ell}] =  \frac{1}{2} \int \rd r \left\{ (\psi'_{\ell})^2 - V_{\ell}(r)\psi_{\ell}^2\right\} \,,
\ee
where we have set aside the label $m$ without loss generality. Indeed, we consider variations and transformations  of the field which act on each mode $m$ separately mixing them.
Let us recall that the equation of motion are given by the Sturm-Louiville equation (\ref{eom}) while the general solution for a general multipole is given by the Legendre functions (\ref{sol}). 

\subsection{HJPSS symmetry from the Sch\"{o}dinger symmetry}
\label{sec41}

As seen in the previous section, this system possesses a set of hidden conserved charges which form a centrally extended Schr\"{o}dinger algebra. Let us write down explicitly the infinitesimal generators and the central charge associated with these conserved charges for the Schwarzschild black hole. Since the symmetry of the multipole $\ell$ can be deduced from the symmetry of the zero mode, i.e. $\ell=0$, by mean of the ladder operators (\ref{ladder}), we focus for simplicity on the $\ell=0$ sector. In that case, we have
\be
  (\psi_{0})_1 = \sqrt{\r2A}  \;, \qquad   (\psi_{0})_2 = \frac{1}{2} \sqrt{\r2A} \log{\left( \frac{r^2}{\r2A}\right)} \,.
\ee
The central charge of the Schr\"odinger algebra is given by the Wronskian of those two solutions, which coincides with the mass of the black hole
\be
\label{c}
w = - \frac{r_s}{2}\,.
\ee
This result echoes the many works on black hole entropy from diffeormorphisms preserving near-horizon boundary conditions, where the resulting symmetry is generated by a Virasoro algebra whose central charge is related to the black hole mass \cite{Iyer:1994ys, Carlip:1995cd, Carlip:1998wz, Cadoni:2005ej, Hotta:2000gx, Carlip:2011vr, Castro:2010fd, Majhi:2012tf}. See also Refs. \cite{Compere:2012jk, Donnay:2016ejv, Akhmedov:2017ftb, Lust:2017gez, Haco:2018ske, Averin:2018owq, Averin:2019zsi, Perry:2020ndy, Grumiller:2019fmp, Chen:2020nyh, Chandrasekaran:2020wwn, Adami:2020amw} for further developments on these near-horizon conformal symmetries. In that context, the central charge then controls the entropy and the ensuing Hawking temperature. Since the scalar field we study is meant to represent (idealized) perturbations of the Schwarzschild metric, it is interesting to wonder if the presently uncovered Schr\"odinger algebra could be interpreted as the truncation, or remnant, of the full Virasoro algebra to scalar perturbations in the static approximation. This conjecture goes well beyond the scope of the present paper and is postponed to future investigation.

\smallskip
\medskip
Now, the generators of the $\sl(2,\mathbb{R})$ symmetry are given by
\begin{align}
\label{gen1}
\delta_{Q_{+}} & = - \r2A \partial_r + \r2A' \,,\\
\label{gen2}
\delta_{Q_{-}} & = - \r2A \log{\left( \frac{r^2}{\r2A}\right)} \partial_r  - \frac{1}{\sqrt{\r2A}} \left[ r_s - \r2A'  \log{\left( \frac{r^2}{\r2A}\right)} \right]\,, \\
\delta_{Q_{0}} & = - \r2A \log{\left( \frac{r^2}{\r2A}\right)}\partial_r  - \frac{1}{2} \left[ r_s - \r2A' \log{\left( \frac{r^2}{\r2A}\right)}   \right]\,,
\end{align}
while the Galilean transformations do not involve differential operators,
\be
\delta_{Y_{+}} = \sqrt{\r2A} \;, \qquad \delta_{Y_{-}} = \sqrt{\r2A}\log{\left( \frac{r^2}{\r2A}\right)}
\,.
\ee
The symmetry generators for an arbitrary $\ell$-multipole can be obtained from them using the ladder operators (\ref{ladder}) as follows
\be
\delta^{\ell}_i  = D^{+}_{\ell-1} .... D^{+}_1 \delta^0_i D^{-}_0 .... D^{-}_{\ell-1} \,.
\ee
At this stage, we can already compare the generators we have obtained with the ones discussed in Ref.~\cite{Hui:2021vcv}. We recognize the generator $\delta_{Q_{+}}$ as the generator of the HJPSS horizontal symmetry for $\ell=0$. Indeed, it leaves invariant the growing mode, i.e. $ (\psi_{0})_1$, i.e. $\delta_{Q_{+}}  (\psi_{0})_1 =0$, while changing the decaying mode, i.e. $\delta_{Q_{+}}  (\psi_{0})_2 \neq 0$. The difference in the explicit form of the generators comes from the difference of rescaling (\ref{rescaling}) used in the present work and in Ref.~\cite{Hui:2021vcv}. We conclude that the horizontal HJPSS symmetry stands as a one-parameter subgroup belonging to the larger Schr\"{o}dinger symmetry group discussed here.

 The role of the other generators is also interesting. In contrast to the HJPSS symmetry, we see that the generator $\delta_{Q_{-}}$ leaves the decaying mode invariant, i.e. $\delta_{Q_{-}} (\psi_{0})_2 =0$, while modifying the growing mode, i.e. $\delta_{Q_{-}} (\psi_{0})_1 \neq 0$. From that point of view, the reasoning developed in Ref.~\cite{Hui:2021vcv} is not fully satisfactory. Rejecting the decaying branch $(\psi_{0})_2$ because it spontaneously breaks the symmetry $\delta_{Q_{+}}$ is not the correct criterion since one can argue the same for the growing branch w.r.t. the symmetry generated by $\delta_{Q_{-}}$. Thus, the question is what is special in the symmetry $\delta_{Q_{+}}$? 
 
 As we shall see, the two other $\sl(2,\mathbb{R})$ symmetries $\delta_{Q_{-}}$ and $\delta_{Q_{0}}$ are spontaneously broken at the horizon. Selecting only the well-defined symmetry reduces the $\SL(2,\mathbb{R})$ group to a one-parameter sub-group generated by $\delta_{Q_{+}}$.

\subsection{Symmetry breaking at the horizon}\label{sec42}

To understand this symmetry breaking, let us come back to the definition of the symmetry. A symmetry maps solutions to the e.o.m. onto other solutions to the same e.o.m. At the level of the action $S[\psi]$, symmetries describe field variations which do not change the bulk value of $S[\psi]$. Thus it might seem that all classical solutions related by a symmetry transformation have the same value of the action. This is a na\"ive point of view and symmetry transformations actually induce boundary terms which contribute to the action variation.
Here, it turns out that if we stretch the domain of definition of the field to the horizon at $r=r_{s}$, some of the  Schr\"odinger symmetry transformations actually produce infinite boundary terms and are not valid symmetry anymore. This leads to a spontaneous symmetry breaking due to the horizon.

\smallskip

In order to discuss more concretely this symmetry breaking, notice that the action for the static linear perturbations is actually playing the role of the potential for the \textit{dynamical} linear perturbations. Consider indeed the dynamical perturbations $\varphi(t,r,\theta, \phi)$ and the background
\be
\rd s^2 = - f(r) \rd t^2 + \frac{\rd r^2}{f(r)} + r^2 \rd \Omega^2
\ee
Rescaling the field as $\psi (t,r,,\theta, \phi)=\sqrt{z} \varphi$ with $z = r^2 f(r)$, the action for the dynamical rescaled perturbation is given by
\be
S[\psi] = \frac{1}{2} \int \rd t \rd r \sin{\theta} \rd \theta \rd \phi \left[ - \frac{1}{f^2} (\partial_t \psi)^2 + \left\{ (\partial_r \psi)^2 - \frac{\mu^2}{z} \psi^2 - \frac{1}{z} \psi \Delta_{S_{2}} \psi \right\} \right]
\ee
Now, focusing on the static problem, i.e. $\partial_t \psi=0$, the profile of the perturbation is dictated by the spatial part of this action which corresponds to the potential of the dynamical perturbation. Indeed proceeding to the Legendre transform, the momenta and the hamiltonian of the dynamical scalar field are given by
\be
\pi = \frac{\delta \L}{\delta \partial_t \psi} \;, \qquad H[\psi] = K[\psi] + V[\psi] = \int \rd r \sin{\theta} \rd \theta \rd \phi \left\{ f^2 \pi^2 + \left[ (\partial_r \psi)^2 - \frac{\mu}{z^2} \psi^2 - \frac{1}{z} \psi \Delta_{S_{2}} \psi\right]\right\}
\ee
Here, $K[\psi]$ is the kinetic energy while $V[\psi]$ plays the role of the potential which coincides with the action for the static scalar field. 
Therefore, the allowed configurations of the static perturbation are given by the minima of the potential $V[\psi]$, i.e :
\be
V_{\text{min}} :=V(\psi_{\text{min}}) < \infty \;, \qquad \frac{\delta V}{\delta \psi} \big{|}_{\psi_{\text{min}}} =0
\ee
Now, considering the symmetry of the action $V[\psi]$ for the static perturbations, the off-shell symmetries transform the shape of the potential, while the on-shell symmetries map the minima of the potential $V[\psi]$ into another one, as expected for a dynamical symmetry. Therefore, our criteria amounts at demanding for two minima $\psi^1_{\text{min}}$ and $\psi^2_{\text{min}}$ related by a symmetry transformation, the transformation of the potential satisfies
\be
\label{delta}
\Delta V_{\text{min}} = V(\psi^2_{\text{min}}) - V(\psi^1_{\text{min}}) = B_{\text{on shell}} < \infty
\ee
where $B$ is the boundary term induced by our symmetry. Now, the symmetry is broken if there is two admissible solutions of the equation of motion which are not related by a symmetry transformation. In our case, this translates into demanding that the different values of the minimum of the potential one can reach with our symmetry remains finite, i.e. (\ref{delta}). This difference is precisely given by the on-shell value of the boundary term induced by the symmetry. Therefore, the criteria we will use in order to select the well defined symmetry transformations is that it does not diverge on-shell. In particular, it can be non-vanishing or zero.

In order to apply this criteria, we now come back to the infinitesimal transformation of the action in order to evaluate the on-shell values of the boundary term.
Let us omit the subscript $\ell$ for simplicity. The $\sl(2,\mathbb{R})$ infinitesimal transformations shift the action by a boundary term explicitly given by
\begin{align}
\label{bd}
\delta_{\xi} S =  \int^{r_{\ast}}_{r_s} \rd r \frac{\rd B}{\rd r}  \;, \qquad \text{with}  \qquad B(r) :=  \left( V \xi + \frac{\xi''}{2}\right) \psi^2
\end{align}
where we integrate from the horizon $r_s$ to some finite radius $r_{\ast}$.
In order for these symmetries to be well-defined, the boundary term has to remain finite in the range $[r_s, r_{\ast}]$. Using (\ref{lam}) and evaluating the boundary on a general solution, i.e. $\psi \simeq c_1 \psi_1+ c_2 \psi_2$, we find that
\be
B(r) \simeq \left[\alpha_{+}\left(  \psi'_1\right)^2  + \alpha_{-}\left(  \psi'_2\right)^2 + \alpha_{0}  \psi'_1\psi'_2 \right] \left( c_1 \psi_1 + c_2 \psi_2 \right)^2 \,.
\ee
At this stage, the divergences arise only from the behavior of the second branch $\psi_2$, its derivative $\psi'_2$ and the potential $V$ at the horizon.
Finiteness of this boundary term then imposes that
\be
\alpha_{-} = \alpha_0 = 0  \;, \qquad \text{and} \qquad c_2 =0 \,.
\ee
From this symmetry criterion, only the one-parameter symmetry transformation labelled by $\alpha_{+}$ survives, which corresponds to the generator $\delta_{Q_{+}}$. This provides the justification to only consider this sub-group of symmetry. Moreover, the symmetry argument also imposes that $c_2=0$, showing that this residual symmetry is only consistent with a purely  growing branch. Let us stress that the symmetry breaking  criterion discussed here is based on the transformation of the action and not the transformation properties of the solution as in Ref.~\cite{Hui:2021vcv}. This allows one to capture the key difference between the different $\sl(2,\mathbb{R})$ transformations on the physical system.

We can proceed the same way with the Galilean symmetry. The infinitesimal shift of the action induced by this transformation reads
\be
\delta_{\chi} S = \int^{r_{\ast}}_{r_s} \rd r \frac{\rd B}{\rd r}  \;, \qquad \text{with}  \qquad B(r) :=  \left( \chi + 2 \psi \right) \chi' \,.
\ee
Evaluating it on-shell, we obtain
\be
B(r) \simeq  (\eta_{+} \psi'_1 + \eta_{-} \psi'_2) \left[ (\eta_{+} + 2 c_1) \psi_1 + (\eta_{-} + 2 c_2) \psi_2\right] \,.
\ee
Imposing finiteness of this term, we recover the condition $c_2=0$ and the additional condition $\eta_{-} =0$. Therefore, only the Galilean symmetry generated by $\delta_{Y_{+}}$ survives this criterion. 
From the above discussion, we conclude that the Schr\"odinger algebra of the system breaks down to the abelian sub-algebra
\begin{align}
\label{algebra}
 \{ Q_{+}, Y_{+}\} = 0 \,.
\end{align}
This algebra encodes the symmetry protecting the physical solution which is not divergent at the horizon. In the 4-dimensional case, this abelian global symmetry also protects the vanishing of the static Love numbers for the Schwarzschild black hole. Using the on-shell values of the conserved charges $(\ref{onshell})$, we obtain the equivalence 
\be
c_2 =0  \qquad  \Longleftrightarrow \qquad Q_{+} = Y_{+} \simeq 0 \,.
\ee
The interesting outcome of this result is that the value of the conserved charged can be computed far away from the black hole. Therefore, the symmetry criterion allows one to check the properties of the field at the horizon solely from the symmetry satisfied in the asymptotic region, as first advocated in Ref.~\cite{Hui:2021vcv}.

Finally, having identified the sub-group which generates the well-defined transformations of the system, we can wonder if we can write explicitly the associated finite transformations. Let us focus on the zero mode. For the symmetry $\delta_{Q_{+}}$ leaving invariant the growing branch, the associated reparametrization function $f(r)$ shall satisfy the following differential equation
\be
\left(\psi_0\right)_{1}\circ f= (f')^{\f12}\,\left(\psi_0\right)_{1}\;, \qquad \text{where}\qquad \left(\psi_0\right)_{1} = \r2A
\ee
and where we have reintroduced the subscript $\ell=0$ to avoid confusion.
The solution is given by
\be
f_{\lambda}(r) = \frac{\lambda r_s r}{(\lambda-1)r + r_s} \,,
\ee
where $\lambda$ is a real constant which parametrizes this symmetry transformation. It can be checked that this function also solves the condition (\ref{cond0}) when restricting the potential $V$ to its expression~(\ref{e.Vl}) in the case $\ell=0$. We can therefore write down explicitly the set of finite off-shell symmetry transformations for the zero mode. They are compactly written as
\begin{align}
\label{conf}
r \rightarrow \tilde{r} &=  \frac{\lambda r_s r}{(\lambda-1)r + r_s}  \; \,, \\
\psi_0 \rightarrow \tilde{\psi}_0(\tilde{r}) &= \frac{\sqrt{\lambda} r_s}{(\lambda-1)r + r_s}  \psi(r) + \eta_{+} \sqrt{r(r-r_s)} \,,
\end{align}
and parameterized  by the two real constants $(\lambda, \eta_{+})$. Notice that in this simple case, the reparame-trization of the radial coordinate (\ref{conf}) is a special conformal transformation which leaves the horizon invariant, i.e. $f_{\lambda}(r_s) =r_s$. Nevertheless, finding the explicit form of the finite symmetry transformations for an arbitrary multipole $\psi_{\ell}$ turns out to be much more complicated as they involve non-linear third order differential equations. This concludes the discussion on the symmetry breaking and the residual abelian global symmetry dictating the profile of the massless static test scalar field on the Schwarzschild black hole.   

Let us point that the present discussion is based on a specific choice for the boundary term of the action (\ref{acmulti}). Indeed, the same Lagrangian with different boundary terms can describe the same bulk behavior with different boundary physics, such that  the choice of boundary term has to be understood as part of the definition of the theory \cite{Harlow:2019yfa, Freidel:2020xyx}.  In the present discussion, the boundary term has been chosen to vanish. Introducing a suitable boundary term to our Lagrangian might allow one to restore the full Schr\"{o}dinger symmetry but one should then provide an interpretation for these new boundary degrees of freedom. A lot of progress has been done in this direction, and it would be interesting to revisit the symmetry-breaking discussed here using the well-developed edge modes machinery \cite{Carrozza:2021sbk}.

\section{Discussion}

\label{sec5}

We have shown that  any static linear perturbations around a Schwarzschild black hole possess an infinite tower of conserved charges. These charges are constructed from the conserved Wronskians (\ref{cons}) which stand as a special property of the Sturm-Liouville equation dictating the dynamics. From this infinite dimensional charge algebra (\ref{infch}), we have shown that one can identify a finite dimensional subset forming a one dimensional centrally extended Schr\"{o}dinger algebra $\sh(1) = \sl(2,\mathbb{R}) \ltimes \mathcal{H}$. These Schr\"{o}dinger charges generate suitable Galilean transformations (\ref{sym22}) and conformal reparametrization of the field and the coordinate, i.e. (\ref{sym1}), which stand as global Noether symmetries for the static linear perturbations. We stress that our analysis is very general as it holds for any $d$-dimensional spherically symmetric background, while it can include massive and higher spins fields (suitably rescaled and decomposed in spin-weighted spherical harmonics).

We have further shown that the rather complicated $\SL(2,\mathbb{R})$ finite symmetry transformation satisfying (\ref{cond0}) can be understood in a straightforward manner using the so-called conformal bridge. This map transforms the system $(x, \psi)$ with a potential $V$ onto the free system $(F, \Phi)$. The new trivializing coordinate $F$ is constructed as the ratio of the two linearly independent solutions for $\psi$, i.e. $F = \psi_2/\psi_1$. In this alternative picture, the $\SL(2,\mathbb{R})$ finite symmetry simply amounts at the freedom to define this new trivializing coordinate up to a standard M\"{o}bius transformation (\ref{mob}). From that perspective, it is always possible to trivialize the static response of the black hole by identifying this trivializing coordinate.  In terms of this privileged coordinate, the hidden symmetry discussed here simply descends from the standard Schr\"{o}dinger symmetry of the free particle. See appendix~\ref{app} for details.

In the second part of this work, we have used this structure to revisit the static response of the Schwarzschild black hole from a symmetry-based perspective. Following Refs.~ \cite{Kol:2011vg, Hui:2020xxx, Charalambous:2021kcz}, we have adopted the test field approximation and consider for simplicity the case of a massless static scalar field on the Schwarzschild black hole. We have derived the infinitesimal generators of the Schr\"{o}dinger symmetry for this system and shown that one of the $\sl(2,\mathbb{R})$ transformation, i.e. (\ref{gen1}), reproduces the horizontal HJPSS symmetry recently presented in Ref.~\cite{Hui:2021vcv}. For each multipole, this symmetry leaves the growing branch of the solution invariant. This result clarifies the origin of this HJPSS symmetry and in particular the associated explicit finite transformation at the level of the action (\ref{acmulti}). Moreover, the generality of our set-up shows that this symmetry also holds for more general spherically symmetric backgrounds as well as for massive test fields. Let us point that in the free field representation, this HJPSS symmetry takes an even simpler form as it amounts at the invariance of the field under translation along the trivializing coordinate (\ref{sl}).

We have further argued that the symmetry criterion used in Ref.~\cite{Hui:2021vcv} has to be improved. Indeed, from the three $\sl(2,\mathbb{R})$ generators, one of them, i.e. (\ref{gen2}), leaves the decaying branch of the solution invariant while the growing branch spontaneously breaks this symmetry. Therefore, the action of the symmetry generators on the solution is not sufficient to select the symmetry protecting the regularity of the field at the horizon. Instead, we have proposed to use the transformation of the action to discriminate the different symmetries. Demanding finiteness of the on-shell boundary term (\ref{bd}) induced by the symmetry transformations at the level of the action provides the right symmetry criterion. It automatically imposes that the only well-defined symmetry transformations are generated by $\delta_{Q_{+}}$, in agreement with the HJPSS result, as well as by $\delta_{Y_{+}}$. As a consequence, the Schr\"{o}dinger symmetry is broken down at the horizon to an abelian symmetry.  This residual global symmetry protects the regularity of each multipole of the static scalar field at the horizon.

Whether one can consider this symmetry as explaining the vanishing of the Love numbers for the 4-dimensional Schwarzschild black hole depends on what we ask. Indeed, a static massless scalar field on a $d$-dimensional Schwarzschild black hole will also possess a Schr\"{o}dinger symmetry. One can again identify the sub-group of symmetry protecting the regularity of the field at the horizon, but in that case, the Love numbers will not be vanishing anymore. The relation between the regularity of the field at the horizon and the vanishing of the Love numbers is a peculiar feature of the dimension 4 case which is not explained by the present symmetry. With this word of caution, it is nevertheless true that the abelian symmetry identified in this work provides a symmetry protection for this specific property of the 4-dimensional Schwarzschild black hole.

At this stage, let us point several open directions to be explored. First, the symmetry criterion we have used depends crucially on our choice of boundary term for the action (\ref{acmulti}), which vanishes in the present case. Whether one can restore the full Schr\"{o}dinger symmetry at the horizon by adding a suitable boundary term and give a consistent interpretation in terms of edge modes is an interesting direction to be pursued. However, a limitation to do so is that the present symmetry works only for a given multipole, such that adding a covariant boundary term to the scalar field action which could restore the symmetry for all the multipoles remains elusive at this point~\footnote{For metric perturbations, a covariant boundary term should be common not only for all multipoles of linear perturbations but also for the background and nonlinear perturbations.}. From a more general point of view, it should be possible to view the present symmetry transformation acting on the static field $\psi_{\ell}(r)$ as the truncation of a more general symmetry acting on the full field $\psi(r, \theta, \phi)$. Such a more complete picture will be discussed elsewhere. 

Another interesting question to be explored is whether we can understand the origin of this Schr\"{o}dinger symmetry for the test fields from the symmetries of the Schwarzschild mechanics. Indeed, we have seen that the explicit finite symmetry transformations for the zero mode, given by (\ref{conf}), stands as special conformal reparametrization of the radial coordinate. A natural question is how the Schwarzschild background transforms under such conformal reparametrization ? It has been shown recently in Ref.~\cite{Achour:2021dtj} that the Schwarzschild-(A)dS mechanics possesses a set of dynamical symmetries under the Poincar\'e group $\SL(2,\mathbb{R})\ltimes \mathbb{R}^3$, extending previous results in Ref.~\cite{Geiller:2020xze}. In particular, the Schwarzschild background was shown to be covariant under the M\"{o}bius symmetry transformation provided the metric functions transform with a suitable conformal weight\footnote{See Eq~(4.2), Eq~(4.4) and Eq~(4.6) in Section~4 in  Ref.~\cite{Achour:2021dtj}}. Therefore, at least for the zero mode, the specific transformation (\ref{conf}) is also a conformal symmetry of the background. What prevents us from generalizing this conclusion is that the symmetry for the higher multipoles are not M\"{o}bius transformations but more complicated reparametrization of the radial coordinate. Once more, finding the more general symmetry acting on the test field $\psi(r, \theta, \phi)$ and not only on each multipole would be desirable, as it might provide the key to understand if this observation hides a deeper connection with the symmetry of black hole mechanics recently discussed in Ref.~\cite{Achour:2021dtj}.

Finally, let us point that the Schr\"{o}dinger symmetry is only broken down at the horizon. Then, the fact that a static linear scalar perturbation on the Schwarzschild black hole possesses a non-relativistic conformal invariance and that the associated central charge (\ref{c}) coincides with the Schwarzschild mass is reminiscent of the many works exhibiting emergent near horizon conformal symmetries for black holes \cite{Iyer:1994ys, Carlip:1995cd, Carlip:1998wz, Cadoni:2005ej, Hotta:2000gx, Carlip:2011vr, Castro:2010fd, Majhi:2012tf, Compere:2012jk, Donnay:2016ejv, Akhmedov:2017ftb, Lust:2017gez, Haco:2018ske, Averin:2018owq, Averin:2019zsi, Perry:2020ndy, Grumiller:2019fmp, Chen:2020nyh, Chandrasekaran:2020wwn, Adami:2020amw}. Whether the current Schr\"{o}dinger algebra  can be understood as some type of truncation of these more general charges algebra associated with diffeomorphism compatible with the horizon suggests an interesting direction to be explored. In particular, it might allows one to connect our results with complementary works such as Ref.~\cite{Charalambous:2021kcz}.

\bigskip

 \medskip

 \textbf{Acknowledgments}\smallskip

 \medskip
The work of J. Ben Achour was supported by the Alexander von Humboldt foundation.
The work of S. Mukohyama was supported by JSPS KAKENHI No.\ 17H02890, No.\ 17H06359, and by World Premier International Research Center Initiative, MEXT, Japan.

\appendix

\section{Sturm-Liouville equation for static perturbations on Kerr}

\label{kerr}

In this appendix, we show that the static linear perturbations around a Kerr black hole, modelled by a massless test scalar field, can be also recast in term of a Sturm-Liouville equation. Moreover, we provide the explicit form of the solutions for the scalar field profile. This shows that the general symmetry discussed in this work and the argument developed above holds for the rotating black hole case.

Consider the Kerr black hole metric
\be
\rd s^2 = - \frac{\Delta}{\rho^2} \left( \rd t - a \sin^2{\theta} \rd \phi \right)^2 + \frac{ \sin^2{\theta}}{\rho^2} \left[ (r^2 + a^2) \rd \phi - a \rd t \right]^2 + \frac{\rho^2}{\Delta} \rd r^2 + \rho^2 \rd \theta^2 
\ee
where the functions $(\Delta, \rho)$ are given by
\begin{align}
\Delta = r^2 - 2Mr + a^2 \;, \qquad \rho^2 = r^2 + a^2 \cos^2{\theta} .
\end{align}
The parameter $M$ encodes the mass of the black hole and $J = a M$ its angular momentum. Consider the action for massless test scalar field given by
\be
S[\varphi] = \frac{1}{2} \int \rd^4x \sqrt{|g|} g^{\mu\nu} \partial_{\mu} \varphi \partial_{\nu} \varphi
\ee
We are interested in the static case, i.e. $\partial_t \varphi =0$. Introducing the rescaled field $\psi = \sqrt{\Delta} \varphi$ and performing a decomposition in spherical harmonics, i.e.
\be
\psi (r, \theta, \phi) = \sum_{\ell, m} \psi_{\ell, m}(r) Y_{\ell, m} (\theta, \phi)  \qquad \text{with} \qquad Y_{\ell,m} := e^{i m \phi} P^m_{\ell} (\cos{\theta})\;,
\ee
the action decomposes into
\be
S[\psi] = \sum_{\ell,m} S_{\ell,m}  \qquad \text{with} \qquad S_{\ell,m} = \frac{1}{2} \int \rd r \left\{ |\partial_r \psi_{\ell,m}|^2 - V_{\ell,m} |\psi_{\ell,m}|^2\right\}
\ee
where the potential reads
\be
\label{pot}
V_{\ell,m}(r) = \frac{M^2 - a^2}{\Delta^2} + \frac{a^2 m^2}{\Delta^2} - \frac{\ell(\ell+1)}{\Delta} 
\ee
It follows that the profile of a given mode $\ell$ of the static linear perturbation satisfies the following Sturm-Liouville equation
\be
\psi''_{\ell,m} + V_{\ell,m} \psi_{\ell,m} =0
\ee
The explicit solution is given by the two branches
\be
\psi_{\ell,m}(r) = \sqrt{\Delta} \left[ c_1 P^{\mu}_{\ell} \left( \frac{r-M}{\sqrt{M^2 - a^2}}\right) + c_2 Q^{\mu}_{\ell} \left( \frac{r-M}{\sqrt{M^2 - a^2}} \right)  \right]
\ee
where the parameter $\mu$ is complex and given by
\be
\mu = \frac{i a m}{\sqrt{M^2-a^2}}
\ee
The two branches correspond to the Legendre functions of the second kind and one can check that the second branch diverges at the outer horizon of the Kerr black hole. Thus, the problem of the static response of the Kerr black hole within the test field approximation can be recast in term of a Sturm-Liouville equation. It follows that the general results presented in Section~\ref{sec3} apply to the static linear perturbations of the Kerr black hole.
\section{Alternative realization of the symmetry}

\label{app}

In this appendix, we present an alternative realization of the Schr\"{o}dinger symmetry for the static linear perturbations of black holes which allows one to connect with the standard Schr\"{o}dinger symmetry of the free particle.

To see this, let us first recall that one can transform the action for a one dimensional field in a inhomogeneous potential into the action of the free particle. For the time-dependent harmonic oscillator, this conformal mapping is the well-known Arnold transformation \cite{Arnold2, Guerrera, QAM, Guerrero:2013bva, Dhasmana:2021qvw}. When the pulsation is a constant, this map reduces to the Niederer transformation \cite{Niederer}. Such conformal bridge between a system with a non-trivial potential and the free particle have been generalized to more complex systems. See Refs.~\cite{Galajinsky:2010ry, Inzunza:2019sct, Inzunza:2021vgt} for details, Refs.~\cite{BenAchour:2020xif, Achour:2021lqq} for recent applications to cosmological systems and Ref.~\cite{Achour:2021dtj} for black hole mechanics. 

\subsection{Conformal bridge and M\"{o}bius symmetry}

Let us briefly review the construction of the conformal bridge. Consider the field redefinition 
\be
\psi(x) = (F')^{-1/2} \Phi \circ F(x) \,,
\ee
where the function $F(x)$ satisfies
\be
\label{cond1}
\text{Sch}[F] = - 2 V(x)\,.
\ee
Upon performing this field redefinition, the action for the self-interacting field $\psi(x)$  (\ref{ac}) takes the form of the action for the free particle up to a boundary term; see Eq.~(\ref{cond0}). The effect of this field redefinition amounts at encoding the physical information on the potential $V(x)$ in the new coordinate $F(x)$. The explicit form of this coordinate is obtained by solving the condition (\ref{cond1}). Given two solutions to the e.o.m., i.e. $\rd^2_x\psi + V \psi \simeq 0$, say $(\psi_1, \psi_2)$, an explicit solution is given by 
\be
F =  \left\{ \frac{\psi_1}{\psi_2}, \frac{\psi_2}{\psi_1}\right\}.
\ee
This coordinate can be understood as trivializing the motion. Indeed, the equation of motion on $\psi$ is equivalent to a free motion for the field $\Phi$:
\be
\pp_{x}^{2}\psi+V\psi\simeq0
\quad\Leftrightarrow\quad
\pp_{F}^{2}\Phi\simeq0
\,.
\ee
This allows one to solve the differential equation in a very simple manner. Indeed the free system equation $\pp_{F}^{2}\Phi=0$ admits two independent solutions, the constant and the linear solutions,
\be
\pp_{F}^{2}\Phi\simeq0 \quad\Leftrightarrow\quad
\Phi(F)\simeq C_1+C_2 F\,,
\ee
where $(C_1,C_2)$ are real constants to be fixed by suitable boundary conditions. They correspond to the two independent solutions of the original field equation:
\be
\begin{array}{lcl}
\Phi_{1}=1 &\Rightarrow& \psi_{1}=(F')^{-\f12} \\
\Phi_{2}=F &\Rightarrow& \psi_{2}=(F')^{-\f12}F\,.
\end{array}
\ee
We can now discuss how the Schr\"{o}dinger symmetry of the self-interacting system is realized in this alternative picture.

Starting with  $\delta_{Q_{+}}$ \,,
we know that the finite conformal reparametrization it generates leaves $\psi_{1}$ invariant. In other words, we are looking for reparametrizations $f(x)$ such that:
\be
\psi_{1}\circ f= (f')^{\f12}\,\psi_{1}\,.
\ee
It turns out that there is a one-parameter family of such functions 
which we identify as translations in the trivializing coordinate $F$. Let us define these translations as $f_{\lambda}$ such that
\be
F(f_{\lambda}(x))=F(x)+\lambda\,.
\ee
Differentiating this relation, we obtain:
\be
f'_{\lambda}\,F'\circ f_{\lambda}=F'\,.
\ee
Remembering that $\psi_{1}$ is expressed in terms of $F$, we easily show that this equation is exactly the same as above:
\beq
\psi_{1}\circ f= (f')^{\f12}\,\psi_{1}
&\Leftrightarrow &
(F')^{-\f12}\circ f = (f')^{\f12}\,(F')^{-\f12}
\nn\\
&\Leftrightarrow &
F'=f'\,F'\circ f
\,.
\eeq
Similarly, exchanging the roles of $\psi_{1}$ and $\psi_{2}$, we can conclude that the finite symmetries generated by  $\delta_{Q_{-}}$ are the translations in $1/F$.
Finally, the symmetry $\delta_{Q_0}$ 
 generates dilatations inversely acting on $\psi_{1}$ and $\psi_{2}$, i.e. functions $f_{\lambda}$ such that:
\be
\psi_{1}\circ f_{\lambda} =\lambda (f')^{\f12}\psi_{1}
\quad\textrm{and}\quad
\psi_{2}\circ f_{\lambda} =\f1\lambda (f')^{\f12}\psi_{2}
\,.
\ee
These are simply identified as dilatations in the trivializing coordinate
\be
F(f_{\lambda}(x))=\lambda F(x)\,.
\ee
We see that the $\SL(2,\R)$ group of symmetry, i.e. of conformal reparametrization stabilizing the potential, is simply the group of M\"obius transformations on the trivializing coordinate $F$, which is generated by translations, dilatations and the inversion of $F$. This symmetry of the free field $\Phi (F)$ reads
\begin{align}
F \rightarrow \tilde{F} & = M(F)  \;, \qquad \text{with} \qquad M(F) = \frac{a F + b}{c F + d} \,,\\
\Phi \rightarrow \tilde{\Phi}(\tilde{F}) & = (M')^{1/2} \Phi(F) \,,
\end{align}
where $(a,b,c,d)$ are real constants satisfying $ad-bc \neq 0$.
This somewhat surprising fact is natural from the point of view of the definition of $F$ as the ratio of two independent classical solutions, which can thus be redefined after arbitrary linear redefinitions of the chosen two solutions (as long as the Wronskian remains the same).

\subsection{Trivializing the static response of black hole}

\label{sec51}

Let us now construct this alternative representation of the dynamics and discuss the Schr\"{o}dinger symmetry in that case.
By construction, the trivializing coordinate is given by
\be
R_{\ell} (r)= \frac{(\psi_{\ell})_1}{(\psi_{\ell})_2} = \frac{Q_{\ell}(2r/r_s-1)}{P_{\ell}(2r/r_s-1)}
\ee
such that there is one such coordinate for each $\ell$-multipole.
Notice that in this new coordinate, the horizon located at $r_s$ is pushed to $R_{\ell} \rightarrow -\infty$.
Now, the dynamics of each multipole is simply given by
\be
\rd^2_{R_{\ell}} \Phi_{\ell} = 0 \;, \qquad \text{where} \qquad \Phi_{\ell} (R_{\ell}) = (R_{\ell}')^{1/2} \psi_{\ell}(r) \,.
\ee
The profile of the $\ell$-multipole reduces to
\be
\Phi_{\ell} (R_{\ell}) = C_1 + C_2 R_{\ell}
\ee
such that the decaying mode which diverges at the horizon corresponds to the linear term while the growing one is given by the constant mode. 
In this representation of the dynamics, the rather complicated form of the $\SL(2,\mathbb{R})$ symmetry simplifies to
\begin{align}
R_{\ell} \rightarrow \tilde{R}_{\ell} & = \frac{\alpha R_{\ell} + \beta}{\gamma R_{\ell} + \delta} \,, \\
\Phi  \rightarrow \tilde{\Phi}(\tilde{R}_{\ell}) & = \frac{\Phi(R_{\ell})}{\gamma R_{\ell} + \delta}\,.
\end{align}
The infinitesimal $\sl(2,\mathbb{R})$ generators are given by
\be
\label{sl}
\delta_{Q_{+}} = - \partial_R \;, \qquad \delta_{Q_{-}} = - R^2 \partial_R + R \;, \qquad \delta_{Q_0} = - R \partial_R + \frac{1}{2} \,.
\ee
We can now compute the explicit shift of the action induced by these $\sl(2,\mathbb{R})$ transformations. They shift the value of the static action by a boundary term. The symmetry criterion amounts at imposing finiteness of this boundary term. The general finite transformation is given by
\be
\tilde{S}[\tilde{\Phi}_{\ell}]  - S[\Phi_{\ell}] = \int^{R_{\ast}}_{-\infty} \rd R \frac{\rd B}{\rd R} \;, \qquad \text{with} \qquad B(R) := \frac{f''}{f'} \Phi^2_{\ell} \,,
\ee
where we integrate from some finite but large radius $R_{\ast}$ down to the horizon. Computing explicitly the boundary term, we obtain
\be
B(R) = \frac{f''}{f'} \Phi^2 = - 2\gamma \frac{(C_1 + C_2 R)^2}{(\gamma R + \delta)} \,.
\ee
When restricting the integration to a bulk region, the boundary term remains finite which implies that the full $\SL(2,\mathbb{R})$ symmetry is realized. Yet, at horizon, when $R \rightarrow -\infty$, the boundary term remains finite only for $C_2=0$. Finiteness of the boundary term thus imposes that the linear mode, which represents the decaying branch in the standard picture, is absent. The symmetry is therefore spontaneously broken by the horizon. The only $\sl(2,\mathbb{R})$ transformation which survives is given by $\delta_{Q_{+}}$ which amounts at a translation in $R$. Finally, the Galilean symmetry reads
\begin{align}
R \rightarrow \tilde{R} & = R \,, \\
\Phi \rightarrow \tilde{\Phi} (\tilde{R}) & = \Phi(R) + \eta_{+} + \eta_{-} R \,,
\end{align}
whose infinitesimal generators are given by $\delta_{Y_{+}} = -1$ and $\delta_{Y_{-}} = -R$. The boundary term generated by this last symmetry reads
\be
B(R) =  \chi' (\chi + 2 \psi) = \eta_{-} \left[ (\eta_{+} +2 C_1) + (\eta_{-} + 2 C_2 )R\right] \,.
\ee
Again, restricting to the a bulk region, this boundary term remains finite everywhere. However, demanding finiteness of this term at the horizon imposes $\eta_{-}=0$. Thus, only $\delta_{Y_{+}}$ survives this symmetry criterion. We recover the result of Section~\ref{sec42} showing that the Schr\"{o}dinger symmetry breaks at the horizon down to an abelian sub-group generated by $(Q_{+}, Y_{+})$.

\end{document}